\documentstyle[epsf]{mn}

\def\be{\begin{equation}}
\def\ee{\end{equation}}
\def\bea{\begin{eqnarray}}
\def\eea{\end{eqnarray}}
\def\mincir{\raise -2.truept\hbox{\rlap{\hbox{$\sim$}}\raise5.truept
\hbox{$<$}\ }}
\def\magcir {\raise -4.truept\hbox{\rlap{\hbox{$\sim$}}\raise5.truept
\hbox{$>$}\ }}

\title[Mixed models constrained with CMB and LSS]
{CMB AND LARGE SCALE STRUCTURE AS A TEST OF MIXED MODELS WITH $n>1$ }

\author[E. Pierpaoli and S. Bonometto]
{ Elena Pierpaoli$^{1}$ and Silvio Bonometto$^{2,3}$\\ \\
$^1$SISSA -- International School for Advanced Studies,
    Via Beirut 2/4, I34013 Trieste, Italy\\
$^2$Department of Physics of the University,
    Via Celoria 16, Milano, Italy\\
$^3$INFN -- Sezione di Milano  }

\date{Accepted 1997 ** **. Received 1997 ** **; in original form 1997 ** **
}

\begin{document}

\maketitle 

\begin{abstract}

We compute CBR anisotropies in mixed models with different
hot components, including neutrinos or volatile HDM arising from the decay 
of heavier particles.
The CBR power spectra of these models exhibit a higher doppler peak than 
CDM, and the discrepancy
is even stronger in volatile models when the decay gives rise also to a 
neutral scalar.
 
CBR experiments, together with Large Scale Structure (LSS)  data,  are
then used to constrain the space parameter of mixed models, when 
values of the primeval spectral index $n > 1$ are also considered.
Even if $n>1$ is allowed, however, LSS alone prescribes that 
 $\Omega_h \mincir 0.30$.

LSS can be fitted by taking simultaneously a low derelativization redshift 
$z_{der}$ (down to $\simeq 600$) and a high $n$, while CBR data from 
baloon--borne experiment cause a severe selection on this part 
of the parameter space.
In fact, while late derelativization and $n>1$ have opposite
effects on the fluctuation spectrum $P(k)$, they sum their action
on the angular spectrum $C_l$. Henceforth $n \magcir 1.3$ seems
excluded by baloon--borne experiment outputs, while a good fit of
almost all CBR and LSS data is found for $\Omega_h$ values
between 0.11 and 0.16, $n \sim 1.1$ and $z_{der} \sim $2000--5000.
A smaller $n$ is allowed, but  $z_{der}$ should never be  smaller than
$\simeq 1200$.


\end{abstract}


\footnotetext[2]{E-mail: pierpa@sissa.it;\\
bonometto@mi.infn.it}

\section{Introduction}

Anisotropies in the cosmic background radiation (CBR) 
are a strong potential source of information on 
the cosmological model.
Unfortunately, anisotropy observations are hard and significant measures 
were obtained only recently.
As a matter of fact, the theory
of CBR anisotropies 
is
well understood (see, e.g., Hu \& Sugiyama 1995 and references therein) and 
public
numerical codes 
allow to calculate the expected anisotropies 
for a wide range of cases (Seljak \& Zaldarriaga 1996).
It is then easy to see that
CBR anisotropies depend on all the ingredients that 
define a cosmological model:
the background metric, the substance mix and the primeval fluctuation spectrum.
Several authors used available codes to predict CBR features
for suitable ranges of model parameters. However, within the
range of models consistent with the inflationary paradigm,
not enough attention, in our opinion, has been devoted yet
to mixed models. Anisotropies expected for them were calculated by
Ma \& Betschinger (1995), De Gasperis et al. (1995), Dodelson et al. (1996), 
but parameter choices were restricted to cases for which anisotropies only
marginally differ from the standard CDM 
case.
Here  we plan to extend the analysis to a 
wider set of mixed models
 including those
for which a greater discrepancy from standard CDM can be expected and,
in particular, models
with primeval spectral index $n > 1$ and late derelativization of the
hot component.
If hot dark matter (HDM) is made of massive $\nu$'s,
there is a precise constraint between its density parameter
($\Omega_h$) and its derelativization redshift ($z_{der}$):
$$
\Omega_h \simeq 0.68 (z_{der}/10^4)(g_\nu/6)
\eqno (1.1)
$$
(see eqs. 2.3--2.4 below; $g_\nu$ is the number of $\nu$ spin states).
Henceforth, in order to have $\Omega_h \magcir 0.10$--0.15, $z_{der}$
cannot be lower than $\sim 2000$, even for $g_\nu = 6$. In order to
have lower $z_{der}$ and/or greater $\Omega_h$, HDM must arise from
the decay products of heavier particles.
In fact, decay products have extra kinetic energy arising from 
mother particle mass energy and therefore have a later $z_{der}$.
Several authors considered such scenario, assuming
the decay of a heavier neutrino into a lighter one (Bond \&
Efstathiou 1991, Dodelson et al. 1994, White et al. 1995, McNally \&
Peacock 1995, Ghizzardi $\&$ Bonometto 1996),
and there are recent attempts of constraining it
using CBR data (Hannestad 1998). However, 
a wider range of models give rise to similar pictures,
$e.g.$, if metastable
supersymmetric particles
decay into lighter ones (Bonometto et al., 1994; Borgani et al. 1996). 
In a number of recent papers HDM arising from decays was called
$volatile$, to stress its capacity to grant a later derelativization,
which weakens its contribution to the formation of inhomogeneities.
%

CBR anisotropies were first detected by the COBE--DMR experiment 
(Smoot et al. 1992). 
The angular scales observed by COBE were 
rather wide
($\simeq 7^o$) and 
allowed to inspect
a part of the spectrum which is almost substance 
independent. Nevertheless, 
COBE 
measurement 
provide the normalization of density fluctuations
out of the horizon, and 
fair constraints on 
the primeval spectral index
$n$ (Bennet et al. 1996). 
More recent baloon--borne and ground based experiments investigated 
CBR fluctuations on scales comparable with the horizon 
scale at recombination 
In the standard CDM model, on these scales, one expects the first doppler peak
and, 
unlike what happens on larger scales, anisotropies are related
both to the spectral index $n$  and to  substance mix. 
In principle, these scale are the best ones 
to test mixed models, as
on even smaller scales ($l \magcir  500$
see below) 
the CBR spectrum can be 
distorted
by other effects, like reionization and lensing.

The degree scale  measurements currently available 
seem to have detected
the doppler peak 
at a fair angular scale, but with an amplitude
higher 
than 
expected in a standard CDM scenario with $n=1$ (Scott et al. 1995, de 
Bernardis et al. 1997).
An analysis of such 
outputs seems to exclude low values of the
density parameter $\Omega=\rho/\rho_{cr}$ (Hancock et al. 1998, Bartlett et 
al. 1998). Here
$\rho $ is the present average density in the Universe and 
$\rho_{cr} = 3H^2/8\pi G $
depends on the value of the Hubble parameter
 $H = h\, 100\, $km$\, $s$^{-1}$Mpc$^{-1}$ 
Furthermore,
Lineweaver (1997) and Lineweaver $\&$ Barbosa (1998) outline
that
observed anisotropies are still too large to
agree with CDM and $\Omega=1$ unless $h \simeq 0.3$.
Alternative causes for such greater fluctuations
can be 
$n > 1$ or a
cosmic substance 
comprising 
a substantial non--CDM component.

A number of large scale structure (LSS) observables are obtainable 
from the linear theory
of fluctuation growing. A wide range of mixed models predict fair
values for them. Tests of non linear features were also performed
through N--body simulations (see e.g. Ghigna et al. (1997) and referencies 
therein). Although some technical aspects
of mixed model simulations are still questionable, one can state that
suitable DM mixtures allow to fit LSS data
from 1 to 100 Mpc, 
almost up to
scales covered by the current CBR experiments, so that a simultaneous 
analysis of CBR anisotropies and LSS allow  complementary tests of the
models.

It ought to be
outlined that $n>1$ and a hot component have a partially 
compensating effect on LSS and, in a previous paper (Bonometto
$\&$ Pierpaoli 1998, hereafter BP; see also Lucchin et al. 1996,
Liddle et al. 1996), we discussed how they could be combined
to obtain a fit with LSS data. On CBR fluctuations, instead, they
add their effects and a quantitative analysis is needed to see
how far one can go from both $n=1$ and pure CDM. Most of this paper is
focused on such analysis and on the tools needed to perform it.

In particular, let us outline that public codes, like CMBFAST,
cannot predict CBR anisotropies for mixed models with a hot component
of non--thermal origin. A part of this work, therefore, required
a suitable improvement of current algorithms.

In this work only models with 
$\Lambda = 0$, $h = 0.5$ and
$\Omega = 1$ are considered. The {\sl cosmic substance} is fixed 
by the partial density parameters: $\Omega_b = \rho_b/\rho_{cr}$ for baryons,
$\Omega_c = \rho_c/\rho_{cr}$ for cold--dark--matter (CDM),
$\Omega_h$ and $z_{der}$ (see below) for HDM.
Here we shall also distinguish between HDM made of massive
$\nu$'s ($neutrino$ models) and HDM arising from heavier particle decay
($volatile$ models). The relation between the nature of HDM and
the amount of sterile massless components (SMLC hereafter) needs also
to be discussed (in standard CDM, SMLC is made by 3 massless $\nu$'s).

Early deviations from homogeneity are described by the spectrum
$$
P_{\Psi} (k) = A_{\Psi} x^3 (x k)^{n-4}
\eqno (1.1)
$$
($x = x_o - x_{rec}$  is the distance from the recombination band,
as $x_o$ is the present horizon radius). Here $k = 2\pi/L$ and
$L$ is the comoving length--scale. 
Models with $1 \leq n \leq 1.4$ were considered.
In Appendix A we review which 
kinds of inflationary models are consistent with such $n$ interval.

Section 2 is dedicated to a brief discussion on the different kinds of 
hot dark matter  that could lead to a mixed dark matter scenario.
In section 3 we analyze 
the CBR spectrum of
such models, distinguishing between the effects due to the SMLC
and those due to 
the actual phase--space
distributions of the  hot particles
and discussing how current algorithms need to be modified
to provide CBR anisotropies for volatile models.
We then perform an analysis of the parameter space: 
models are preselected according to LSS 
constraints related to the
linear theory. This selection is based on BP results, whose
criteria will be briefly reviewed (section 4). 
BP results, however, were restricted to
the case $\Omega_b = 0.1$. Here we shall inspect a greater portion of
the parameter space, by considering also models with $\Omega_b = 0.05$, 
allowing for a substantial dependence of the CBR power spectrum on the 
baryon abundance.
Section 5 is dedicated to a comparison of the CBR spectra with current
available data, and to the final discussion.

\section{Dark Matter Mix}
 

The {\sl substance} of mixed models can be classified according to their 
behaviour when galactic scales ($ 10^{8}$--$10^{12} M_\odot$)
enter the horizon. Particles already non--relativistic are said 
{\sl cold}. Their individual masses or energy distributions do not
 affect cosmological 
observables. {\sl Tepid} or {\sl warm} components (see, $e.g.$ Pierpaoli et 
al., 1998) become non--relativistic while galactic scales enter the horizon.
{\sl Hot} component(s), instead, become non--relativistic after the latter 
scale has entered the horizon. Neutrinos, if massive, are a typical hot 
component. They were coupled to radiation for $T > T_{\nu,dg} \sim 900\, $keV.
If their mass $m \ll T_{\nu,dg}$, their number density, at any $T < 
T_{\nu,dg}$, is $n_\nu = (3\zeta(3)/4\pi^2) g_\nu T_\nu^3$
(after electron annihilation $T_\nu = T_{\nu,dg} [a_{\nu,dg}/a(t)] 
= (4/11)^{1/3} T$, where $T$ is radiation temperature) and 
their momentum distribution (normalized to unity) reads
$$
\Phi_\nu (p,t) = {2 \over 3 \zeta(3)}
{(p^2/T_\nu^3) \over \exp[p/T_\nu(t)] + 1}
\eqno (2.1)
$$
also when $p \ll m$. Henceforth, when $T \ll m$, their distribution
is not {\sl thermal}, although its shape was originated in thermal equilibrium.
Notice that, for high $p$, $\Phi_\nu$ is cut off as $\exp(-p/T_\nu)$.

Using the distribution (2.1) we can evaluate
$$
\langle p \rangle = (7 \pi^4 / 180 \zeta(3) ) T_\nu = 3.152\, T_\nu~.
\eqno (2.2)
$$
If we define $z_{der} $ as the redshift for which
$\langle p \rangle = m$, eq.~(2.2) tells us that $z_{der}$ occurs when
$T_\nu = 0.317\, m$. In the following we shall use the parameter
$d = 10^4/z_{der}$,
which normalizes $z_{der}$ at a value ($10^4$) in its expected range.
At $z=10^4$, photons and CDM would have an equal density
in a pure CDM model with $h=0.5$ and a present CBR temperature
$T_o = 2.7333\, $K. Henceforth, such redshift is in the
range where we expect that relativistic and non--relativistic
components have equal density (equivalence redshift: $z_{eq}$); 
besides of photons and SMLC, it is possible that HDM contributes 
to the relativistic component at $z_{eq}$. Its value, in different
models, is given by eq.~(2.11), herebelow. In general, we shall
normalize $T_o$ at the above value (which is well inside 1~$\sigma$
in respect to data) and define $\theta = T_o/2.7333\, {\rm K}$.

For neutrinos of mass $m$, 
$$
d = 5.2972 \cdot 4h^2/(m/{\rm eV})~,
\eqno (2.3)
$$
while
$$
\Omega_h = 2.1437 \cdot 10^{-2} g_\nu 
(m/{\rm eV}) \theta^3/4h^2  ~.
\eqno (2.4)
$$

Let us now compare these features with those of a volatile model, where the
hot component
originates 
in the decay
of heavier particles $X$ (mass $m_X$), decoupled since some 
early time $t_{dg}$ (see also Pierpaoli \& Bonometto 1995, and Pierpaoli et 
al. 1997).
If the temperature $T_{dg} \ll m_X$,
such hot component may have a number density much smaller than massive
neutrinos.
Let $N_{X,dg}$ be the heavy particle 
comoving number density 
at decoupling. At $t \gg t_{dg}$ their comoving number density reads:
$$
N_X (t) = N_{X,dg} \exp[-(t-t_{dg})/\tau_{dy}]
\eqno (2.5)
$$
with $t_{dg} \ll \tau_{dy}$ (decay time). Assuming a two--body
decay process $X \to v + \phi$, into a light ({\sl volatile}) particle $v$ 
(mass $m \ll m_X$) and a massless particle $\phi$, it is shown that
the volatile distribution, at $t \gg \tau_{dy}$, reads
$$
\Phi_v (p,t) = 2 (Q/p) \exp(-Q) ~,
\eqno (2.6)
$$
where 
$$
Q = p^2/ {\tilde p}^2
~ {\rm and} ~~~
{\tilde p} = (m_X/2) [a_{dy}/a(t)]
\eqno (2.7)
$$
provided that $X$'s, before they decay, never attain a density exceeding 
relativistic components causing a temporary matter dominated expansion.

At high $p$, the distribution (2.6)
is cutoff $\propto \exp[-(p^2/{\tilde p}^2)]$
and this is true also if this temporary regime occurs.

In BP it is shown that,
if the massless particle $\phi$ is a photon 
($\gamma$), such temporary expansion can never occur in physically relevant
cases, which must however satisfy the restriction $\Omega_h d \ll 1 $.
This limitation does not hold
if $\phi$ is a massless
scalar  as is expected to exist in theories where a 
global invariance is broken below a suitable energy scale (examples of
such particles are {\sl familons} and {\sl majorons}). 

Using the distribution (2.6), it is easy to see that the average
$$
\langle p \rangle = (\sqrt{\pi} / 4) m_X a_{dy}/a(t) =  
{\tilde p} \sqrt{\pi} / 2
\eqno (2.8)
$$
and $v$'s will therefore become non relativistic
when ${\tilde p} = 2\, m/\sqrt{\pi}$; henceforth
$$
{\tilde p} = 2 \cdot 10^4 zmd/\sqrt{\pi}
\eqno (2.9)
$$
can be used in the distribution (2.6), instead of eq.~(2.7).

If $\phi$'s are sterile scalars and the decay takes place
after BBNS, 
they will contribute to SMLC and affect CBR and LSS just as extra
massless neutrino states.

Let us recall that, in the absence of $X$ decay, the ratio
$\rho_\nu / \rho_\gamma = 0.68132(g_\nu/6) \equiv w_o$. $X$ decay 
modifies it, turning $g_\nu$ into an effective value
$$
g_{\nu,eff} = g_\nu + (16/7)(11/4)^{4/3}\Omega_h d ~
\eqno (2.10)
$$
In particular, $\phi$'s lower the {\sl equivalence} 
redshift. For $ d(1-\Omega_h) > 1+w_o$, also $v$'s are still relativistic
at equivalence. Accordingly, the equivalence occurs at either
$$
{\bar z_{eq}} = {4h^{2} \over \theta^4}
{10^4 \over 1+w_o+\Omega_h d}  ~~~~ {\rm or} ~~~~
{\bar z_{eq}} = {4h^{2} \over \theta^4} 
{10^4 (1-\Omega_h) \over 1+w_o+2\Omega_h d}
\eqno (2.11)
$$
in the former and latter case, respectively.

Volatile
models,
 as well as 
neutrino models,
can be parametrized through the values of $\Omega_h$ and $d$. However,
at given $\Omega_h$, the latter ones are allowed only for discrete
$d$ values (notice that such $d$ values are independent of $h$ and
can be only marginally shifted by changing $\theta$).
The former ones, instead, are allowed for a continuous set
of $d$ values. This can be seen in fig.~1, which is taken from
BP, where more details on volatile models can be found. In fig.~1
we also show which models are consistent with LSS constraints 
and COBE quadrupole data,
for various $n > 1$. Such constraints will be 
briefly discussed
in the next section. 
In general, they are fulfilled for a part of the allowed $Q$ values.
Fig.~1 also shows that there is a large deal of mixed models
with low $z_{der}$ which are allowed by LSS data and
are not consistent with HDM made of massive $\nu$'s.
%

\begin{figure}
\centering
\centerline{\mbox{\epsfysize=10.0truecm\epsffile{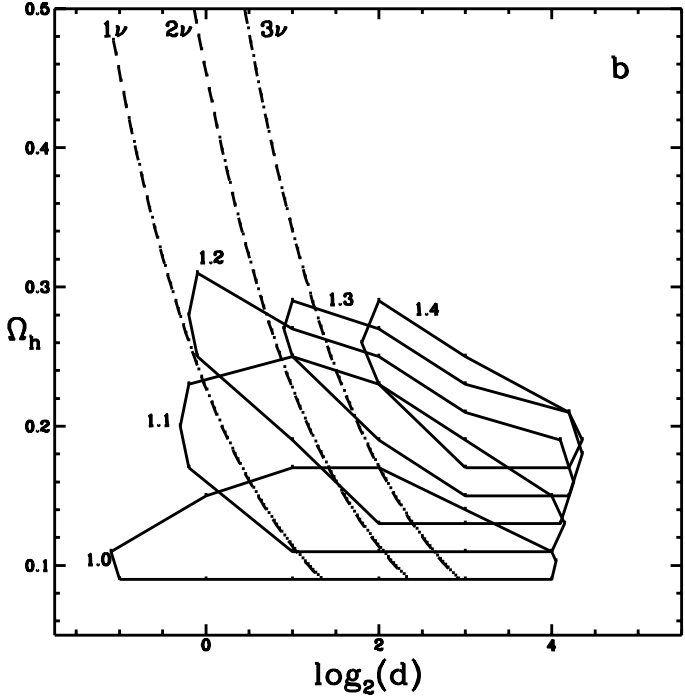}}}
\caption{\it  Parameter constraints from LSS. Dashed lines refer 
 to neutrino cases with 1, 2 and 3 massive neutrinos (from left to right).
 Solid lines are the contours of the regions allowed in volatile models 
 for different values of $n$ (marked next to the areas).
 The plot  refers to $\Omega_b=0.1$.
  The constraints considered are the shape parameter $\Gamma$, 
   the $\sigma_8$ and 
  the number density of clusters are considered (see BP 
  for a detailed discussion).}
\label{fig:rhovsa}
\end{figure}

In this article we shall show that a 
portion of this extra parameter space seems 
however forbidden by CBR constraints.

\section{The radiation power spectrum}

To describe the evolution of radiation anisotropies in an expanding
Universe, it is convenient to write the metric in the form
$$
ds^2 = a^2 (\tau) [ (1+2\Psi) d\tau^2 - (1-2\Phi) \gamma_{\alpha \beta} 
dx^\alpha dx^\beta ] ~.
\eqno (3.1)
$$
in the conformal newtonian gauge. Here $x^\alpha$ (components of 
the vector $\vec{x}$) are space coordinates, $\tau$ is 
the conformal time, $a(\tau)$ is the scale factor and $\gamma_{\alpha\beta}$
gives the spatial part of the metric tensor in the homogeneity
limit. The deviation from a pure Friedmann metric, due to gravitational field
inhomogeneities, are given by the {\it potentials} $\Psi$ and $\Phi$.

In the presence of inhomogeneities, the temperature of radiation 
$T(\hat{n}) = \langle T \rangle [1 + \Delta_T 
(\hat{n})]$ contains an anisotropy term, which can be thought
as a superposition of plain waves of wave--numbers $\vec{k}$. In respect to 
a given direction $\hat{n}$, the amplitude of the single $\vec{k}$ mode 
can be expanded into spherical harmonics. For our statistical aims
it is however sufficient to consider the anisotropy as a function
of $\mu = \hat{k} \cdot \hat{n}$ and use the expansion:
$$
\Delta_T (\vec{k},\hat{n},\tau) = \sum_{l=0}^{l=\infty} (-i)^l(2l+1)
\Delta_l (\vec{k},\tau) P_l(\mu)~,
\eqno (3.2)
$$
where $P_l$ are Legendre polynomials and
whose coefficients can be used to work out the angular fluctuation spectrum
$$
C_l = (4\pi)^2 \int dk\, k^2 P_\psi(k) |\Delta_l(\vec{k},\tau)|^2
\eqno (3.3)
$$
which, for a gaussian random field, completely describes angular anisotropies.

At the present time $\tau_o$ and for a comoving scale given by 
the wavenumber $k$, we can compute $\Delta_l$ performing
a time integral (Seljak and Zaldarriaga, 1996)
$$
\Delta_l(k, \tau_o) = \int_0^{\tau_o} dx  S (k,\tau) j_l [(\tau_o - \tau)k]
\eqno (3.4)
$$
over the source function $S$, which depends upon inhomogeneity evolution
inside the last scattering band and from it to now.

The physics of microwave background anisotropies due to adiabatic 
perturbations has been deeply investigated in the last few years.

It has been shown that the characteristics of the peaks in the $C_l$ 
spectrum
are related to the physics of  acoustic oscillations of baryons and 
radiation between the entry of a scale in the horizon and the last 
scattering band, and on the history of photons from last
scattering surface to us.

Background features, like  the overall matter and radiation  
density content, $h$ and $\Lambda$,
have an influence both on the positions of the peaks and on their 
amplitude,  but the latter  also depends greatly on the baryon
content $\Omega_b$ and more slightly on the characteristic of the hot 
component.

In the following we 
shall analyze in detail the 
angular spectrum of volatile models,
outlining its peculiarities 
with respect to standard CDM and 
neutrino models.
In order to do so, we need to modify available public codes, like CMBFAST,
allowing them to deal with a hot component whose momentum distribution
is (2.6). 

It should also be recalled that
volatile and neutrino models, for given $\Omega_h$ and $d$, 
are expected to include a
different amount of SMLC.
In neutrino models SMLC is less than in pure CDM and even vanishes
if all $\nu$'s are massive (unless extra SMLC is added $ad~hoc$).
In volatile models, instead, SMLC is however more than in pure CDM,
as scalar $\phi$'s are added on top of standard massless $\nu$'s.

Several  $C_l$ spectra of volatile models are presented in figs. 6--15.
They show two main features, if compared with standard CDM: 
  the first  doppler peak is  higher and  the second and 
third doppler peaks are slightly shifted to the right.

In principle, we expect volatile model 
spectra to differ from 
neutrino model spectra because of the momentum distribution of volatiles
and the extra SMLC they have to include. In
the following,  we 
shall
try to disentangle these two effects.

To this aim we coupled each volatile models with a
$technical$ 
neutrino case 
with identical $\Omega_h$ and $d$, but a greater
number of neutrino degrees of freedom,
so to 
ensure equal high-redshift energy densities.
In fig.~2 we report the scale--factor dependence of
the energy densities $\rho (a)$ of volatiles and a massive neutrinos
in two coupled models.
In the case shown, 
the two energy densities never differ in ratio 
more than 
$10^{-3}$;
for
different choice of the parameters
the curve is just
shifted to higher or lower 
redshifts according to 
the value of
$z_{der}$.
\begin{figure}
\centering
\centerline{\mbox{\epsfysize=10.0truecm\epsffile{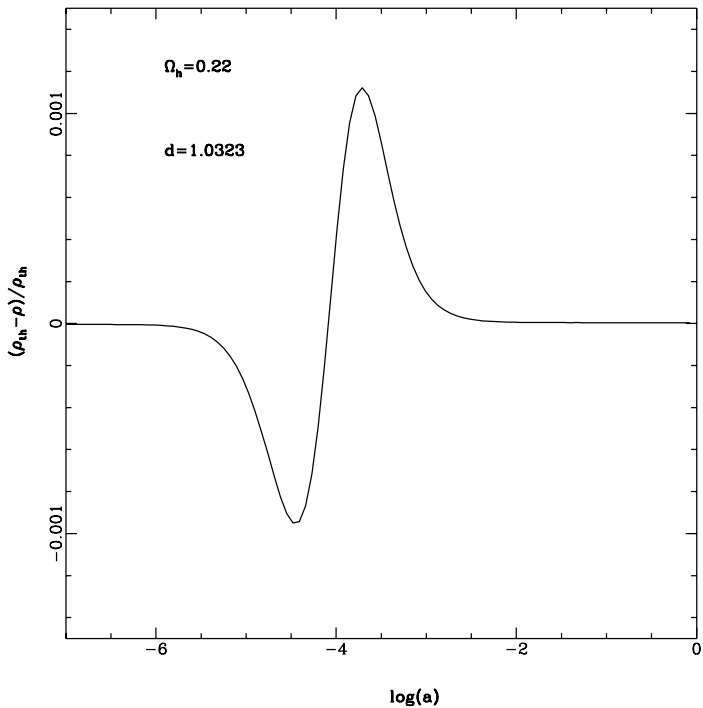}}}
\caption{\it density evolution of a volatile model  in comparison
  with the corresponding  technical neutrino case.}
\label{fig:rhovsa}
\end{figure}

More in detail
fig.~2 states that 
volatiles
have a 
slower
derelativization 
than
neutrinos: 
the transition phase from the relativistic to the  non--relativistic
regime starts earlier and goes on for a longer time.
This behaviour 
is related to
the different shapes of  the two 
distribution functions, 
and to the fact that
the volatile one is 
smoother around 
$\langle p \rangle$, which corresponds to a value significantly
smaller than
its maximum,
after which it is rapidly cutoff [see eqs. (2.6), (2.7)].

Friedman equations show that $\tau(a) \sqrt{\rho (a)}$
is approximately constant.
Hence, once we know $\rho (a)$, we can perform
a comparison 
between the conformal times of coupled volatile and 
$technical$ neutrino cases.
It shows
a marginal discrepancy 
as
already the  $\rho(a)$ in the volatile and
$technical$ neutrino cases are very similar, and moreover 
 the hot component 
always contributes as a 
small
fraction of the total energy density.
On the contrary, if a similar comparison is performed between standard CDM 
and volatile models, big discrepancies are found, especially at high 
redshifts.
In fact, in the volatile cases the relativistic background is greater due 
to the contribution of the sterile component, and
the conformal time is therefore smaller than in the CDM case (see fig.~3).
This  implies visible effects on the position of the doppler peaks,
which are due to the oscillatory phase with which the  photon--baryon fluid 
meets the last scattering band (see Hu \& Sugiyama 1995).
The  photon--baryon fluid oscillates as $cos(k r_s)$, 
where $k$ is the comoving scale and $r_s$ is the sound horizon
($r_s=\int_0^{\tau(a)} {d\tau c_s(a)}$, $c_s(a)$ is the sound speed).
Given the photon--baryon ratio,  $r_s(a)$ follow a similar trend as
$\tau (a)$.
Since in volatile  models $\tau (a)$ is smaller than in CDM, so will be
 $r_s(a)$, and the peaks of the spectrum  will appear in correspondence to
higher $k$ ( i.e. higher $l$) values. 

\begin{figure}
\centering
\centerline{\mbox{\epsfysize=10.0truecm\epsffile{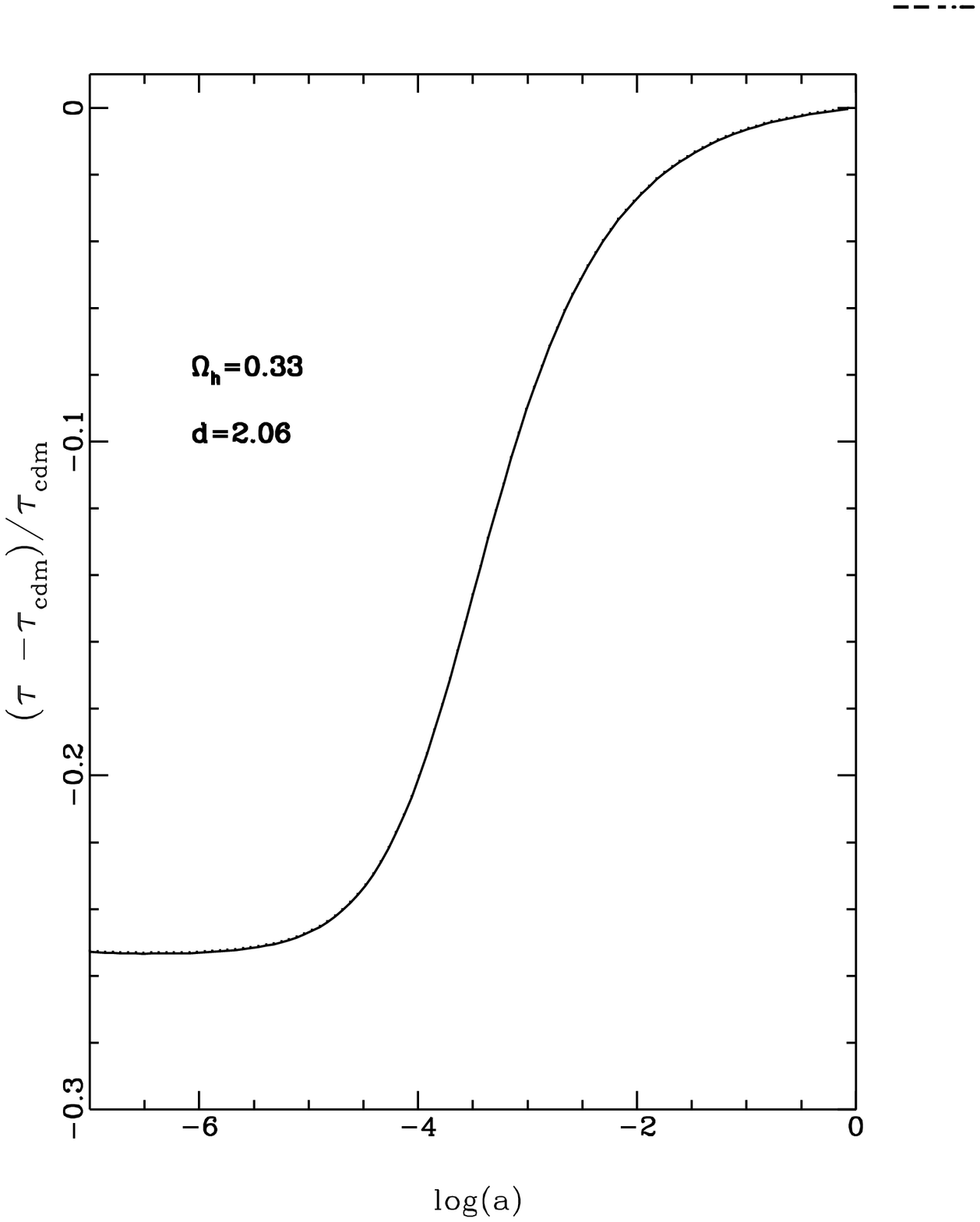}}}
\caption{\it comparison of  conformal time in a CDM standard model and
          in  a volatile model. At high redshifts, volatile models typically 
show a smaller  conformal time.}
\end{figure}

This is a specific features of these models, in  neutrino models the same 
effect plays a role, but shifting the peaks in the opposite direction
(Dodelson et al. 1996). 


For a given $n$,
the height of the peaks is 
fixed
by 
(i) the ratio between baryon and photon
densities, $i.e.$
 $\Omega_b h^2$,
and (ii) the ratio between matter and radiation
densities. At fixed
$\Omega_b$ and $h$ 
the main reason for a higher doppler peak in volatile models
(with respect to CDM) is the delayed matter--radiation
equivalence, for which 
both SMLC and, possibly, volatiles can be
responsible.
In neutrino models without $ad$--$hoc$ SMLC, only the possible
delay due to late derelativizing $\nu$'s may exist.
This is 
why volatile $C_l$ spectra and standard
neutrino ones look so different.
However, there is a tiny 
further contribution in the boost of the peak due to the free-streaming of
the hot component.
 Several authors (Ma \&  Bertschinger 1995, Dodelson et al. 1996) have
shown that even in 
high $z_{der}$ neutrino models
the doppler
peaks are enhanced with respect to CDM, and in that case the
free-streaming of the hot component is to be considered responsible for
the enhancement.
Free--streaming, in fact, causes a decay in the potential $\Phi$ which
contributes as a forcing factor (trough $\ddot \Phi$) in the equations
whose solution are the {\it sonic}
oscillations in the photon--baryon fluid, displacing their zero--point
and, henceforth, the phase by which they enter the last scattering band.
In the standard neutrino case, this effect
causes a variation of $10 \% $ at most on the $C_l$, and typically of
$2 \% $ on the first doppler peak.

In principle one can expect that the different momentum distribution of 
volatiles may alterate the
free--streaming behaviour.
Such differences, if they exist, can be found by comparing
volatile spectra with  the $technical$ neutrino ones.
The differences between the two spectra are presented
are shown
 in fig.~4, and amount to $2 \%$ at most.
Although modest, this is another feature that characterizes volatile models 
with respect to neutrino one.
\begin{figure}
\centering
\centerline{\mbox{\epsfysize=10.0truecm\epsffile{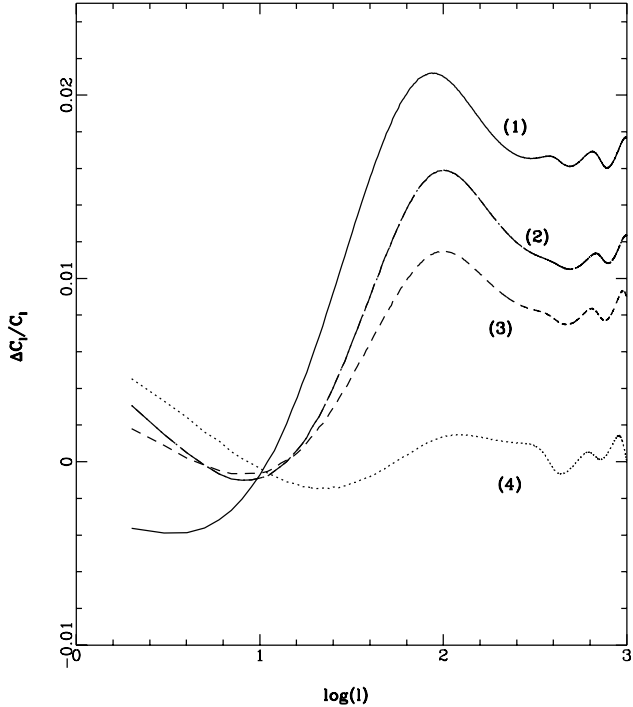}}}
\caption{\it differences in the radiation power spectrum between 
  the volatile case and the techical neutrino case.
  The model parameters are the following: 1) $\Omega_h=0.16, ~~ d=4.25$;
  2)  $\Omega_h=0.22,~~ d=1.03$; 3)$\Omega_h=0.22,~~ d=2.06$;
  4) $\Omega_h=0.33,~~d=2.06$.  }
\label{fig:cldfig}
\end{figure}

In comparison with such finely tuned predictions from theoretical models,
currently available data are still affected by huge errorbars.
However, some feature seems already evident from them. In fig.~6--15
we perform a
%
 comparison of model predictions with 
data and show that the
doppler peak 
observed by the  Saskatoon experiment
(Netterfield et al. 1997) exceeds the one expected in pure
CDM
once it is normalized to
COBE data (Bennet et al. 1996).
%
While it is evident that
volatile models show 
a higher doppler peak, it is clear that a fit
could be reached also changing
other parameters, 
$e.g.$, by
taking $n>1$.
In fig. 4 we show what happens in neutrino  models if the 
spectrum is anti--tilted to $n=1.1$ and to $n=1.2$.
Indeed, the first doppler peak is raised (which is desirable), but 
 also the following peaks are raised, 
making difficult the agreement with 
the results from the CAT experiment (Scott et al. 1996).
\begin{figure}
\centering
\centerline{\mbox{\epsfysize=10.0truecm\epsffile{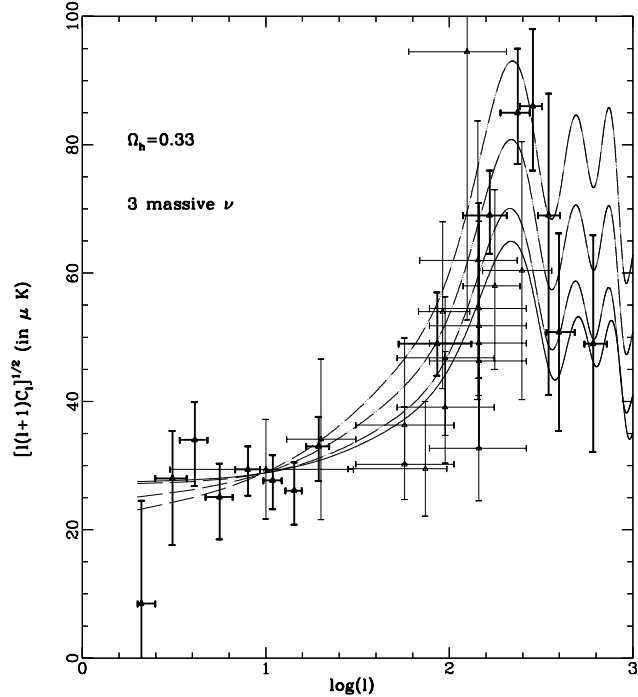}}}
\caption{\it a standard neutrino model with different n values. $C_l$
curves 
from bottom to top correspond to $n = 1, 1.1, 1.2, 1,3$. 
Bold data points refer to 
COBE (Tegmark 1996), 
CAT (Scott et al 1996), Saskatoon (Netterfield et al. 1997); see Scott et 
al (1995) for a summary of all the experiments.}
\label{fig:redhis}
\end{figure}
In section 5 similar considerations will be used in order to constrain the
whole set of volatile models.


\section{Constraints from Large Scale Structure}

Mixed model parameters can be constrained from particle physics 
and/or from LSS. In this section we review a number of the latter constraints,
which can be tested without discussing non--linear evolution. In Appendix
A we debate constraints on the spectral index $n$ arising from inflation.

Even without considering their non--linear evolution, models can be
constrained through the following prescriptions:

(i) The numerical constant $A_{\Psi}$, in the spectrum (1.1), must give
a value of $C_2$ consistent with the COBE quadrupole $Q_{rms, PS}$.
Values of $A_{\Psi}$ consistent with the $Q_{rms, PS}$ values, for a given
$n$, within 3$\, \sigma$'s, can be kept.

(ii) COBE quadrupole therefore fixes the normalization at small $k$. The first
large $k$ test to consider, then, is the behaviour on the $8\, h^{-1}$Mpc
scale. Quite in general, the mass $M_\lambda$, within a sphere of 
radius $L= \lambda h^{-1}{\rm Mpc}$, is
$$
M_\lambda = 
5.96 \cdot 10^{14} \Omega h^2 M_\odot (\lambda/8)^3~.
\eqno (4.1)
$$
Therefore, the $8\, h^{-1}$Mpc scale is a typical cluster scale.
Here optical and X--ray data are to be exploited to work out the
mass variance $\sigma_8$ and models should fit such observational outputs.
Optical data provide the cluster mass function through a virial
analysis of galaxy velocities within clusters. X--ray determinations,
instead, are based on observational temperature functions.
If clusters are substantially virialized and the intracluster 
gas is isothermal, the mass $M$ of a cluster can then 
be obtained, 
once the ratio $\beta_{th,gal} $ between thermal
or galaxy kinetic energy (per unit mass) and gravitational potential
energy (per unit mass) is known. Values for $\beta$'s are currently 
obtained from numerical models.

Henry $\&$ Arnaud (1991) compiled a complete X--ray flux--limited sample of 
25 clusters which is still in use for such determinations. Assuming 
an isothermal gas, full virialization and $\beta
 \equiv \beta_{gal}/\beta_{th}
= 1.2$ they had estimated $\sigma_8 = 0.59 \pm 0.02$. Their error does not
include $\beta$ uncertainty. Various authors then followed analogous patterns 
(see, e.g., White et al. 1993, Viana $\&$ Liddle 1996). Recently Eke et al. 
(1996) used Navarro et al. (1995) cluster simulations to take $\beta = 1$ with
an error $\mincir 6\, \%$. Accordingly they found $\sigma_8 = 0.50 \pm 0.04$.
By comparing the above results one can estimate that, to obtain $\sigma_8 \sim
0.7$, under the assumption of full virialization and purely
isothermal gas, $\beta \sim 1.4$ is needed. 

An estimate of 
cluster masses 
independent from cluster models can be 
obtained by comparing optical and X--ray data. Recent analyses (Girardi
et al. 1998) seem to indicate values of $\beta \simeq 0.88$. In our opinion,
such outputs do not strengthen the case of a safe cluster mass
determination, as they are more than 12 \% below Navarro et al.
(1995) ratio and might indicate a non equilibrium situation.
Furthermore, it ought to be outlined that cluster mass determinations 
based on a pure virial equilibrium assumption
conflict with the observed baryon abundances and would require
cosmological models with $\Omega_b \sim 0.16$--0.20, in contrast with 
BBNS constraints, if all dark matter is CDM and $\Omega = 1$.

If 
HDM
 is only partially bound in clusters and their masses are 
underestimated by $\sim 15$--20$\, \%$, the latter conflict can be
overcome.
(Alternative way outs, of course, are that $\Omega < 1$ or
$\Lambda \neq 0$.) Therefore, in order that data be consistent with mixed 
models, some mechanism should cause a slight but systematic underestimate of
cluster masses.
Owing to such uncertainties, we can state that cluster data constrain 
$\sigma_8$ within the interval 0.46--0.70.

These constraints can also be expressed with direct reference to
the cumulative cluster number density. Defining the mass
$M(u)$ for which the top-hat mass variance $\sigma_{M(u)} = \delta_c/u$ (here
$\delta_c$ values from 1.55 to 1.69 can be considered) the Press $\&$ 
Schechter approach yields the number density
$$
n(>M) = \sqrt{2/\pi} (\rho/M) \int_{\delta_c/\sigma_M}^\infty
du [M/M(u)] \exp(-u^2/2) ~.
\eqno (4.2)
$$
A usual way to compare it with data amounts to taking $M = 4.2h^{-1} \cdot 
10^{14}$M$_\odot$ and considering then  $N_{cl} = n(>M) (100h^{-1} 
{\rm Mpc})^3$ for the above $M$ value. With a range of uncertainty
comparable with the one discussed for $\sigma_8$, optical and X--ray data
converge towards a value of $N_{cl} \sim 4$. Henceforth viable models
should have $1 \mincir N_{cl} \mincir 10$, for one of the above
values of $\delta_c$.

There is a slight difference between testing a model in respect to
$\sigma_8$ or $N_{cl}$. This amounts to the different impact that
the slope of the transferred spectrum has on expected values. 
Observations, however, also constrain the observed spectral slope, as
we shall detail at the point (iv).

(iii) In order to have $N_{cl}$ and $\sigma_8$ consistent with observations,
the $A_{\Psi}$ interval obtained from COBE quadrupole may have to be
restricted. The residual range of $A_\Psi$ values can then be used 
to evaluate the expected density of high--$z$ objects, that mixed models
risk to {\sl under}--produce. The most restrictive constraints comes
from computing $\Omega_{\rm gas} = \alpha \Omega_b \Omega_{\rm coll}$ 
in damped Lyman $\alpha$ systems (for a review  see Wolfe, 1993).

It can be shown that
$$
\Omega_{\rm coll} = {\rm erfc}[\delta_c/\sqrt{2} \sigma(M,z)],
\eqno (4.3)
$$
where $\sigma(M,z) $ is the (top hat) mass variance (for mass $M$ at 
redshift $z$) and $\alpha$ is an efficiency parameter which should be 
$\mincir 1$. More specifically, using such expression, one can evaluate
$D_{Las}(M,z) \equiv \Omega_{\rm gas} \times 10^{3}  /\alpha $. Then,
taking $z = 4.25$, $\delta_c = 1.69$ and $M=5\cdot 10^{9} h^{-1}M_\odot$
we have a figure to compare with the observational value given by
Storrie--Lombardi et al. (1995): $D_{Las} = 2.2 \pm 0.6$. 

Only models for which the predicted value of $D_{Las} $ exceeds 0.5,
at least for a part of the allowed $A_\Psi$ interval, are therefore viable.
In turn, also for viable models, this may yield a further restriction
on the $A_\Psi$ interval.

(iv) Models viable in respect to previous criteria should also have a
fair slope of the transferred spectrum. Its slope can be quantified through
the {\sl extra--power} parameter $\Gamma = 7.13 \cdot 10^{-3}(\sigma_8/
\sigma_{25})^{10/3}$ ($\sigma_{8,25}$ are mass variances on the scales 
$R = 8,25\, h^{-1}$Mpc). Using APM and Abell/ACO samples Peacock and 
Dodds (1994) and  Borgani et al. (1997) obtained $\Gamma $ in the
intervals 0.19--0.27 and 0.18--0.25, respectively. Such intervals essentially
correspond to $2\, \sigma$'s. Furthermore the lower limit can be 
particularly sensitive to underestimates of non--linear effects.
Henceforth, models yielding $\Gamma$ outside an interval 0.13--0.27 are
hardly viable.

One can also test models against bulk velocities reconstructed POTENT from 
observational data. This causes no constraint, at the 2$\, \sigma$ level, 
on models which survived previous tests.

In BP a number of plots of the transferred spectra of viable models,
were shown against LCRS reconstructed spectral points (Lin et al. 1996).
However, previous constraints include most quantitative limitations
and models passing them fit spectral data. Fig.~1 is taken from BP and
reports the curves  on the $\Omega,d$ plane limiting areas where viable 
mixed models 
exist for various primeval $n$ values, if $\Omega_b = 0.1$. All models
considered in the next sections, both for $\Omega_b = 0.1$ and $\Omega_b
= 0.05$, were previously found to satisfy the above constraints.

\section{CMB data and parameter space limitation}

In this section we  give   the CBR spectra of the 
hot--volatile models and compare them with  available data, ranging
from $l=2$ to $l \mincir 500$.

We evaluated the spectra for several parameter choices
allowed by LSS constraints (see fig.~1).
Significant example of $C_l$ spectra are shown 
 in figs.6--15 while the corresponding LSS
predictions are summarized in table 1.


\begin{table}[h]
\centering
\caption[]{Model parameters and power spectra. Column 1:
volatile fractional density; Column 2: redshift
at which the  volatile component  becomes non-relativistic  
($d=10^{4}/z_{der}$);
Column 3: total number of equivalent massless neutrinos ($n_\nu = g_{\nu.
eff}/2$);
Column 4: n value considered;
Columns 5--7: Large scale structure predictions.}
\tabcolsep 6pt
\begin{tabular}{ccccccc}  \\ \hline \hline
${\bf \Omega_h}$ & $d$ & $n$ & $N_{\nu}$ &  ${\bf \sigma_8}$
& ${\bf \Gamma}$ & $N(>M)$ \\ \hline \\
\multicolumn{7}{c}{\bf $\Omega_b=0.05$}\\
\hline
0.11 & 8 & 1 & 6.9  & 0.53--0.68 & 0.23  &  1.8--9.8\\
0.11 & 8 & 1.1 & 6.9 & 0.63 & 0.27 &  6.9 \\
0.16 & 2.83 & 1 & 5 & 0.60--0.68 & 0.20 & 4.0--9.3 \\
0.16 & 2.83 & 1.1 & 5 & 0.67 & 0.24 &  8.9 \\
0.11 & 16 & 1.1 & 10.7 & 0.52--0.64 & 0.24 &  1.6--7.3 \\
0.19 & 8 & 1.1 & 9.7 & 0.55--0.66 & 0.16 &  2.0--7.7 \\
0.19 & 8 & 1.2 & 9.7 & 0.55--0.67 & 0.19 &  2.1--8.1 \\
0.20 & 4 & 1.2 & 6.5 & 0.67 & 0.20 &  8.3 \\
0.24 & 4 & 1.3 & 7.2 & 0.64--0.69 & 0.17 & 6.4--9.4  \\
0.23 & 8 & 1.3 & 8.9 & 0.57--0.68 & 0.17 & 2.7--9.3  \\

\hline
\multicolumn{7}{c}{\bf $\Omega_b=0.1$} \\
\hline
0.11 & 8 & 1 & 6.9  & 0.52--0.65 & 0.21  &  1.4--7.6\\
0.11 & 8 & 1.1 & 6.9 & 0.58--0.67 & 0.24 &  3.6--8.9 \\
0.16 & 2.83 & 1 & 5 & 0.66--0.68 & 0.18 & 7.4--9.0 \\
0.16 & 2.83 & 1.1 & 5 & 0.60--0.65 & 0.21 &  4.4--7.1 \\
0.11 & 16 & 1.1 & 10.7 & 0.47--0.67 & 0.22 &  1.0--7.9 \\
0.19 & 8 & 1.1 & 9.7 & 0.57--0.68 & 0.14 &  2.9--8.7 \\
0.19 & 8 & 1.2 & 9.7 & 0.49--0.67 & 0.17 &  1.0--8.0 \\
0.20 & 4 & 1.2 & 6.5 & 0.66--0.68 & 0.13 &  7.1--8.0 \\
0.24 & 4 & 1.3 & 7.2 & 0.54--0.69 & 0.14 & 1.6--8.8  \\
0.23 & 8 & 1.3 & 8.9 & 0.52--0.70 & 0.15 & 1.2--10  \\
\hline
\end{tabular}
\end{table}

Some parameter sets are compatible with  neutrino hot dark 
matter, while models with a low $z_{der}$
and $n \magcir  1$ are obtainable with volatile hot dark matter. 
Since the height of the first doppler peak is very sensitive to 
the baryon abundance, 
we considered two values of $\Omega_b$, namely 0.05 and 0.1.
Spectra are normalized to $Q_{rms,PS}$ assuming  
no contribution of gravitational waves.
As is known, their contribution would  raise the  low--l tail of the
 $C_l$ spectrum, therefore reducing the gap between the Sacks--Wolfe {\it 
plateau} and the top of the first doppler peak.

Models with $\Omega_b=0.1$ systematically show a peak less pronounced 
than models with  $\Omega_b=0.05$.
It is well known that models with a given $h$ and hot component show a 
lower doppler peak for smaller  $\Omega_b$; in top of that, here there is a 
further effect: LSS constraints often are compatible with a part of the 
observational $Q_{rms,PS}$ interval,  and low $\Omega_b$ models tend to be 
consistent with low $Q_{rms,PS}$ values.

\begin{figure}
\centering
\centerline{\mbox{\epsfysize=10.0truecm\epsffile{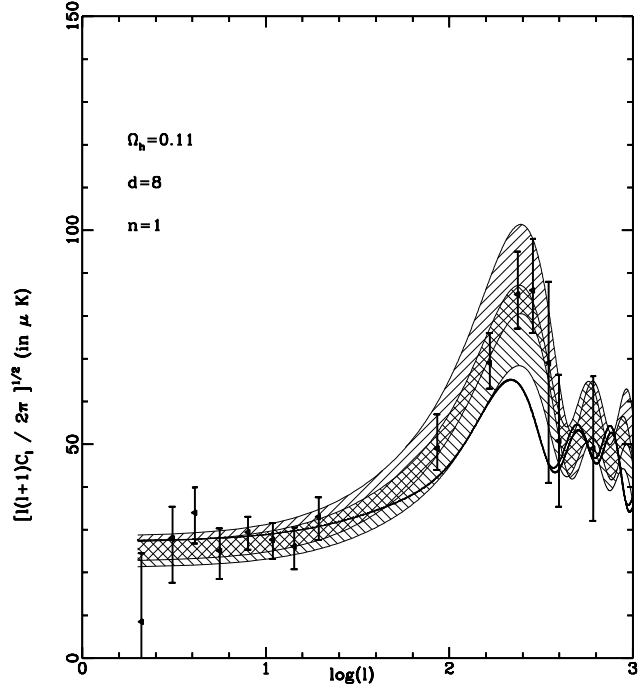}}}
\caption{\it  volatile spectra are compared with the observational 
data from COBE, Saskatoon and CAT experiments. The errorbars
 correspond to 1 $\sigma$ errors.
 Solid bold lines refer to standard CDM ($h=0.5$, $n=1$, $\Omega=1$) and
 a neutrino model with  $\Omega_h=0.3$.
 Shaded areas are the volatile spectra with parameters specified in the 
figure. Upword shading refers to $\Omega=0.1$ models while downword
shading to $\Omega=0.05$
Notice that $z_{der}=10^4/d=1250$.}

\end{figure}

\begin{figure}
\centering
\centerline{\mbox{\epsfysize=10.0truecm\epsffile{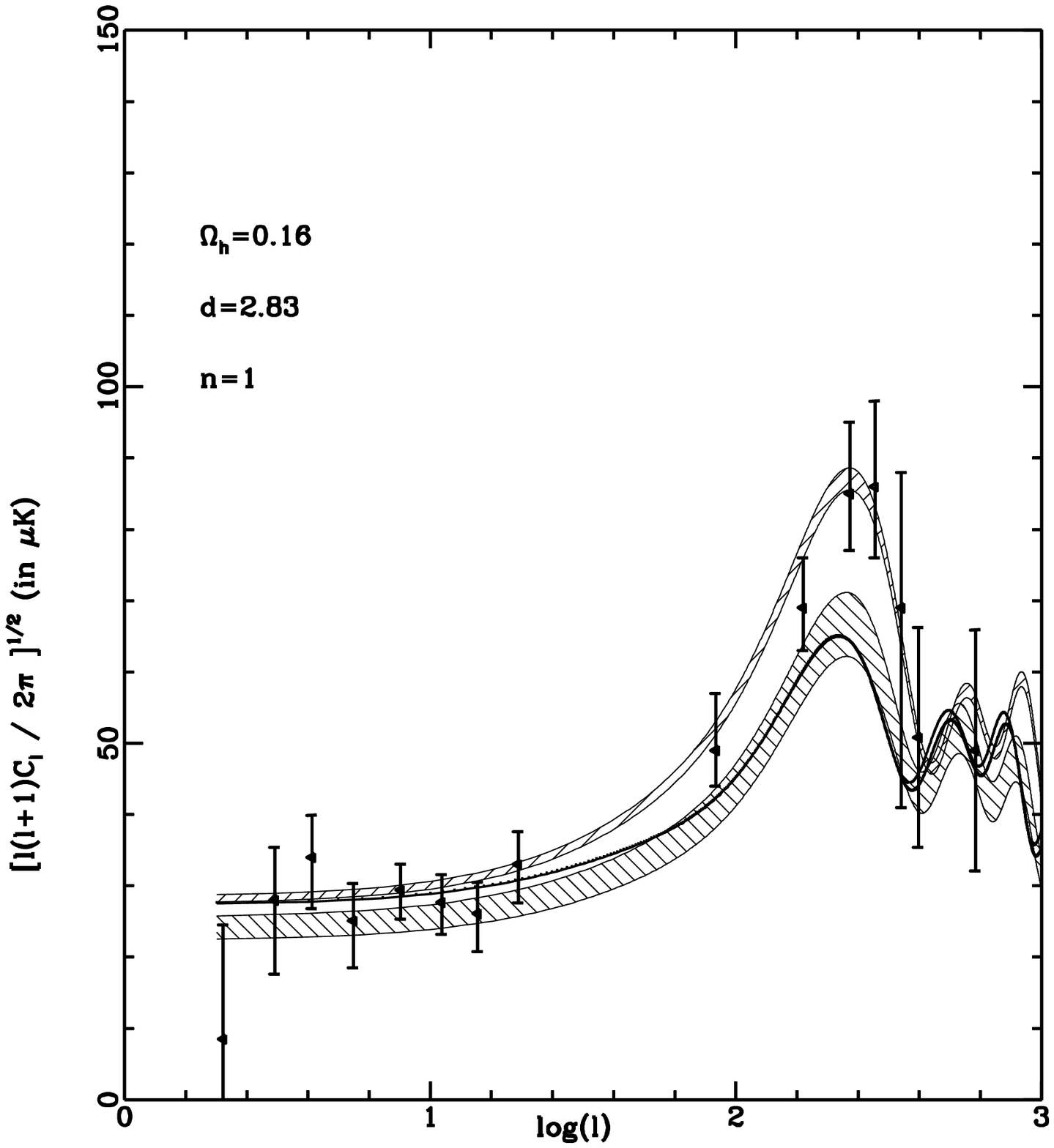}}}
\caption{\it volatile spectra compared to the data. $z_{der}=3533$.}
\end{figure}

\begin{figure}
\centering
\centerline{\mbox{\epsfysize=10.0truecm\epsffile{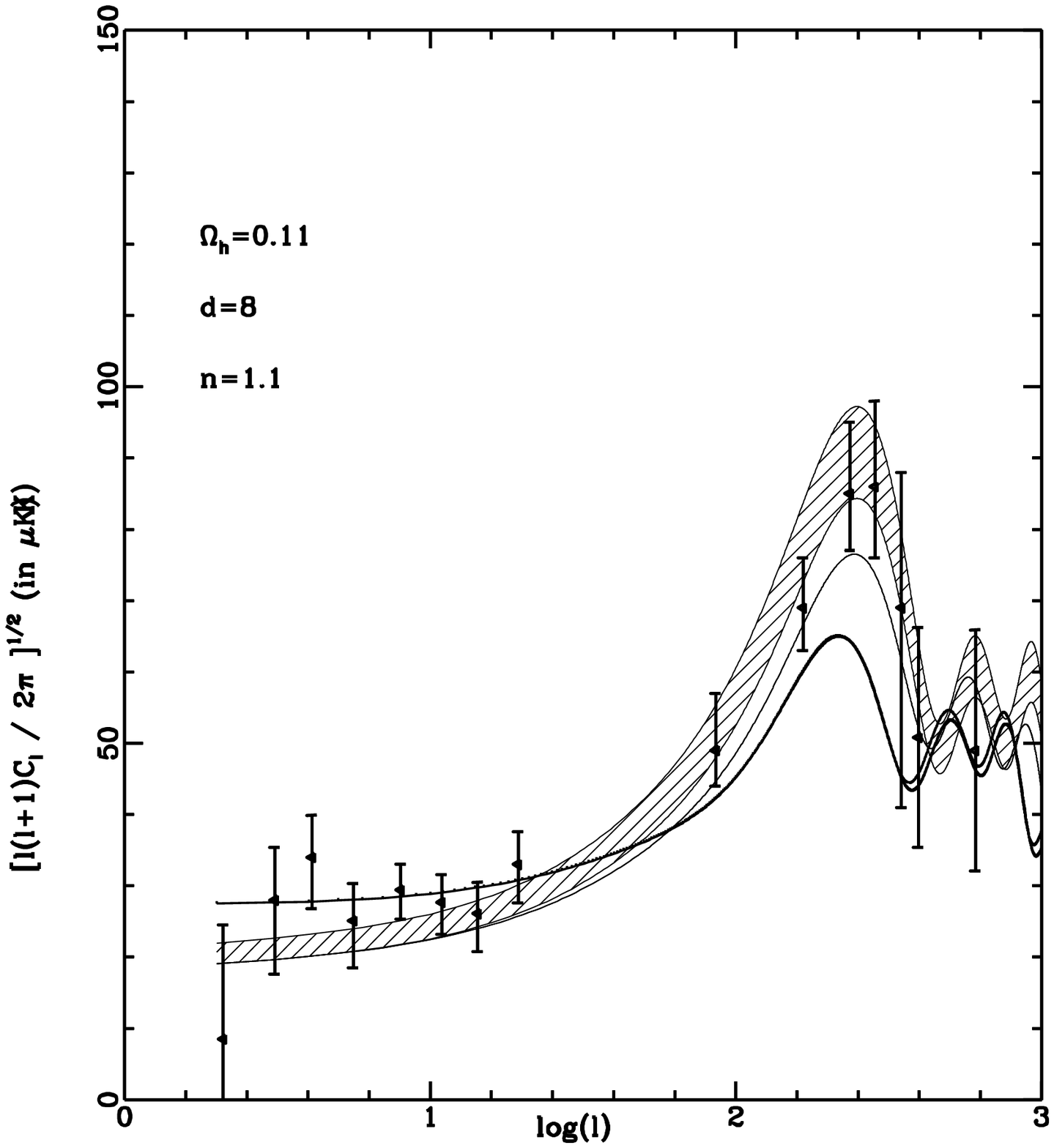}}}
\caption{\it volatile spectra compared to the data. $z_{der}=1250$.}
\end{figure}

\begin{figure}
\centering
\centerline{\mbox{\epsfysize=10.0truecm\epsffile{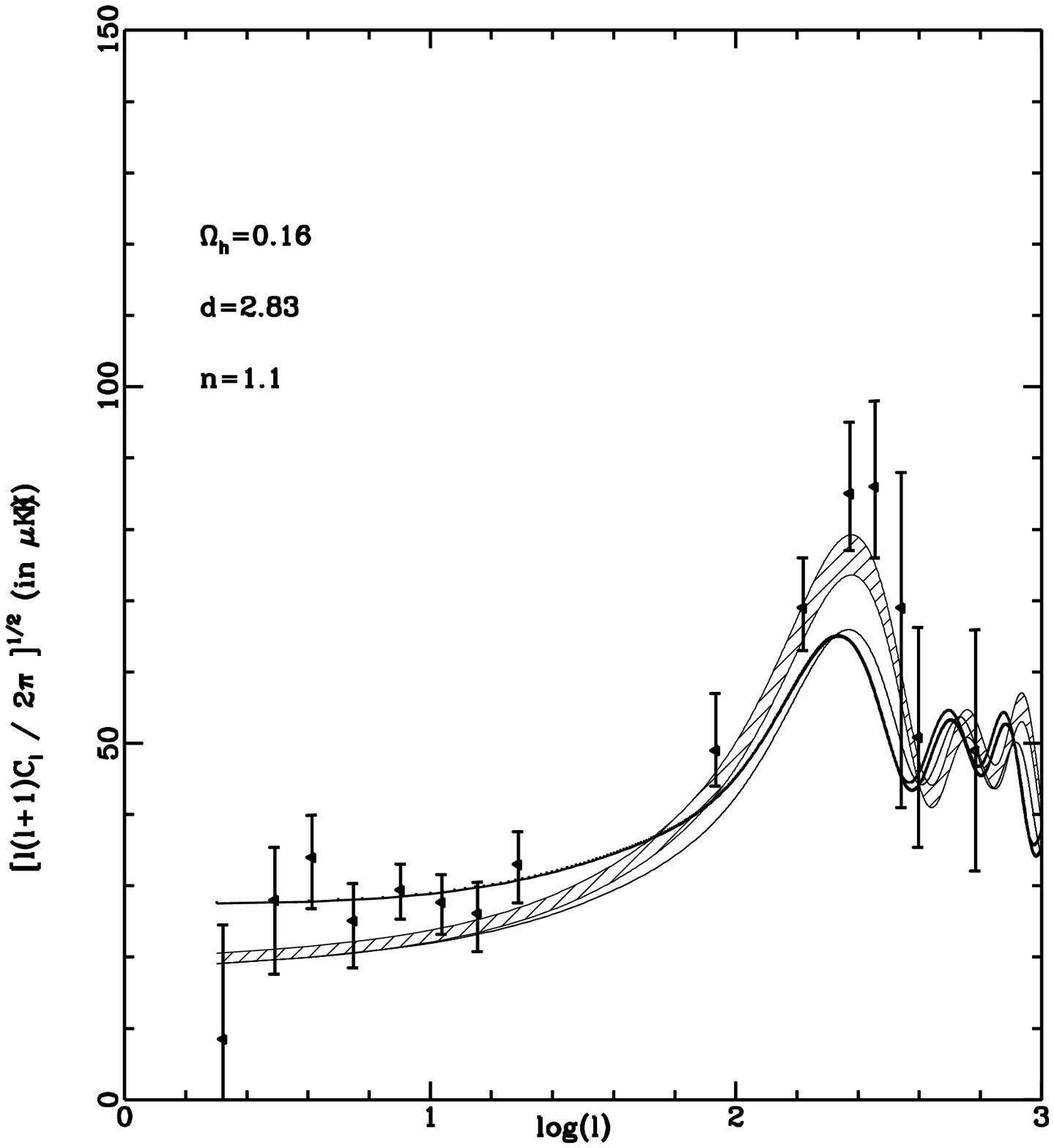}}}
\caption{\it volatile spectra compared to the data.  $z_{der}=3533$.}
\end{figure}

\begin{figure}
\centering
\centerline{\mbox{\epsfysize=10.0truecm\epsffile{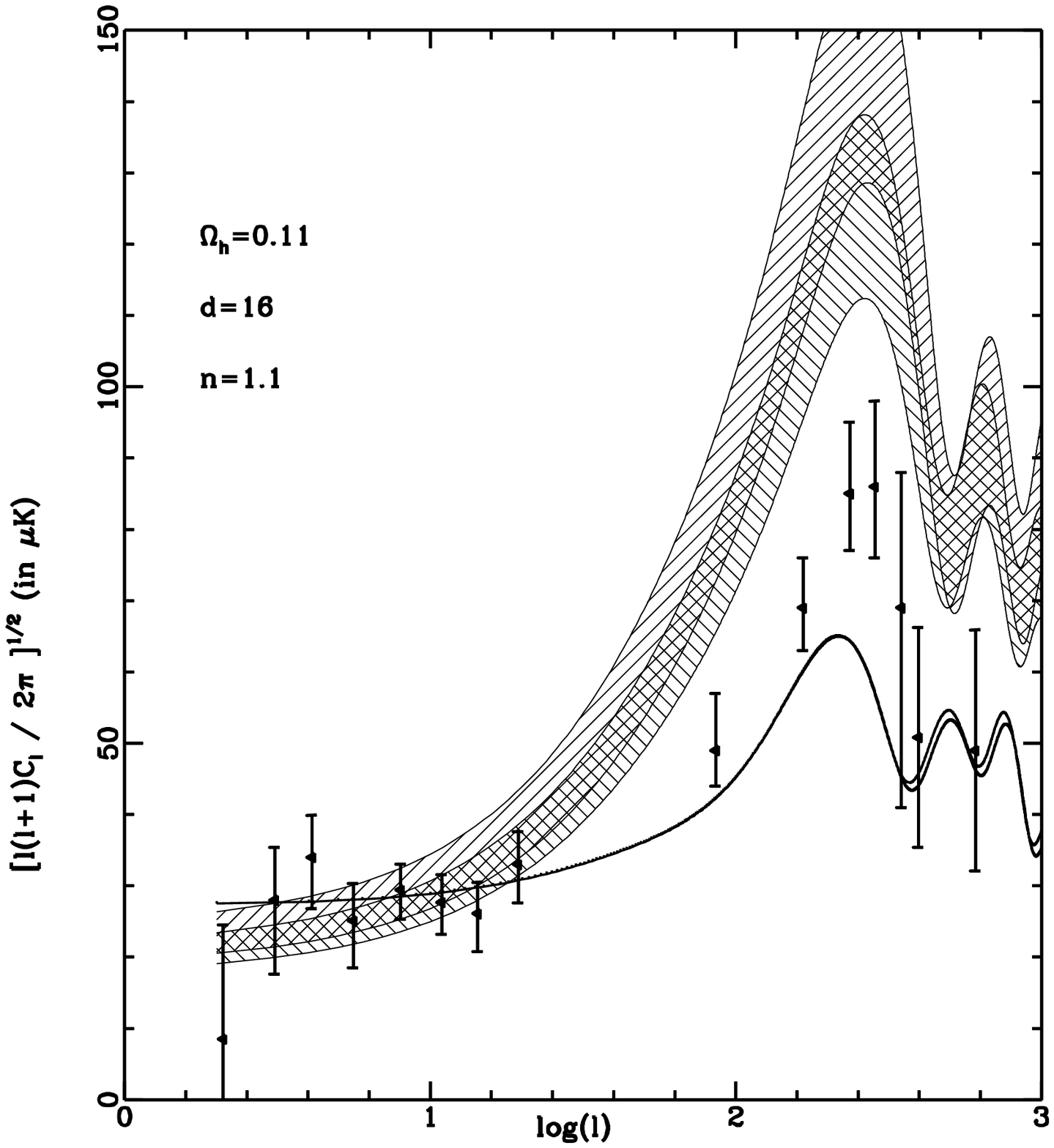}}}
\caption{\it volatile spectra compared to the data. $z_{der}=625$.}
\end{figure}

\begin{figure}
\centering
\centerline{\mbox{\epsfysize=10.0truecm\epsffile{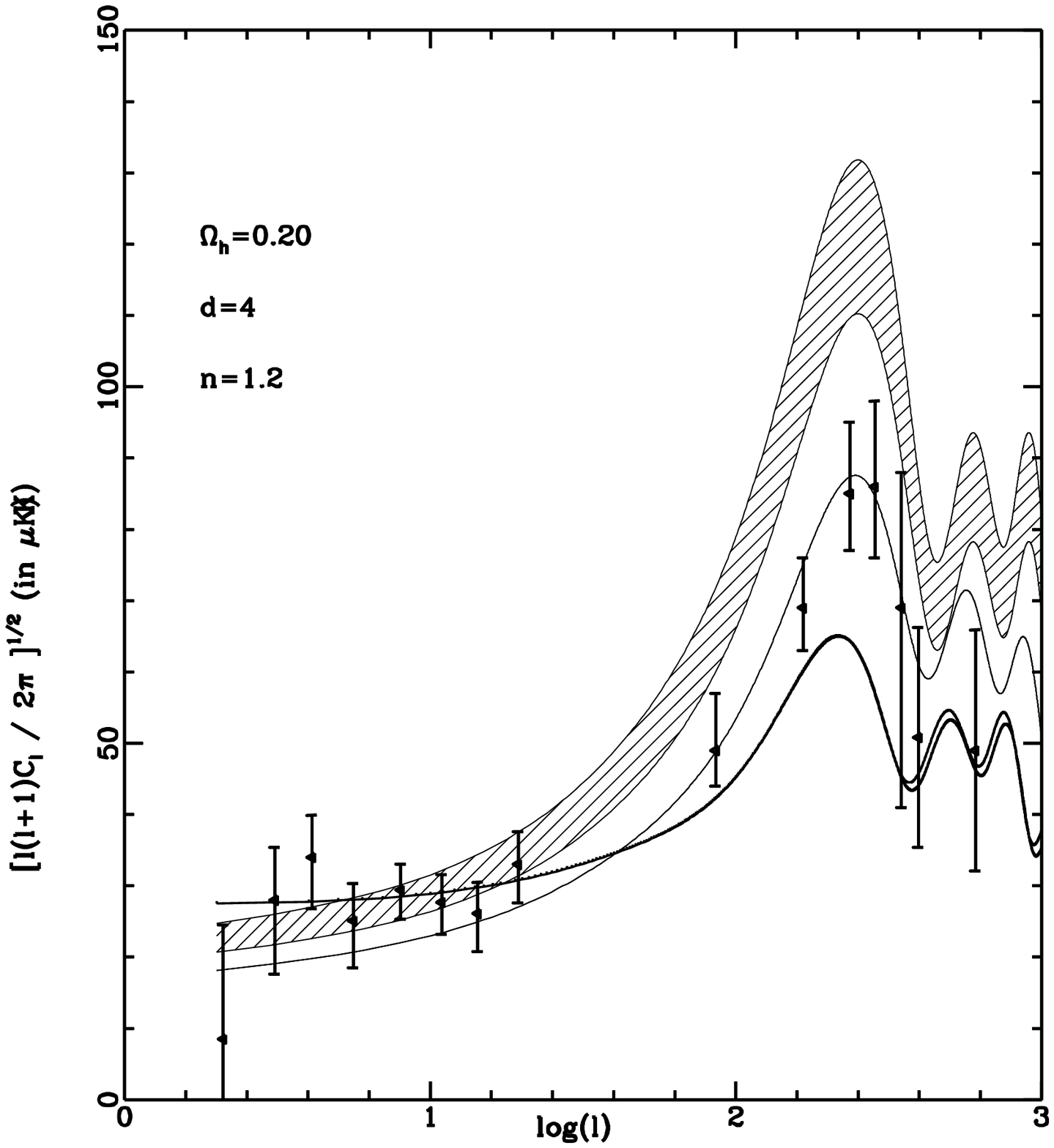}}}
\caption{\it volatile spectra compared to the data.$z_{der}=2500$. }
\end{figure}

\begin{figure}
\centering
\centerline{\mbox{\epsfysize=10.0truecm\epsffile{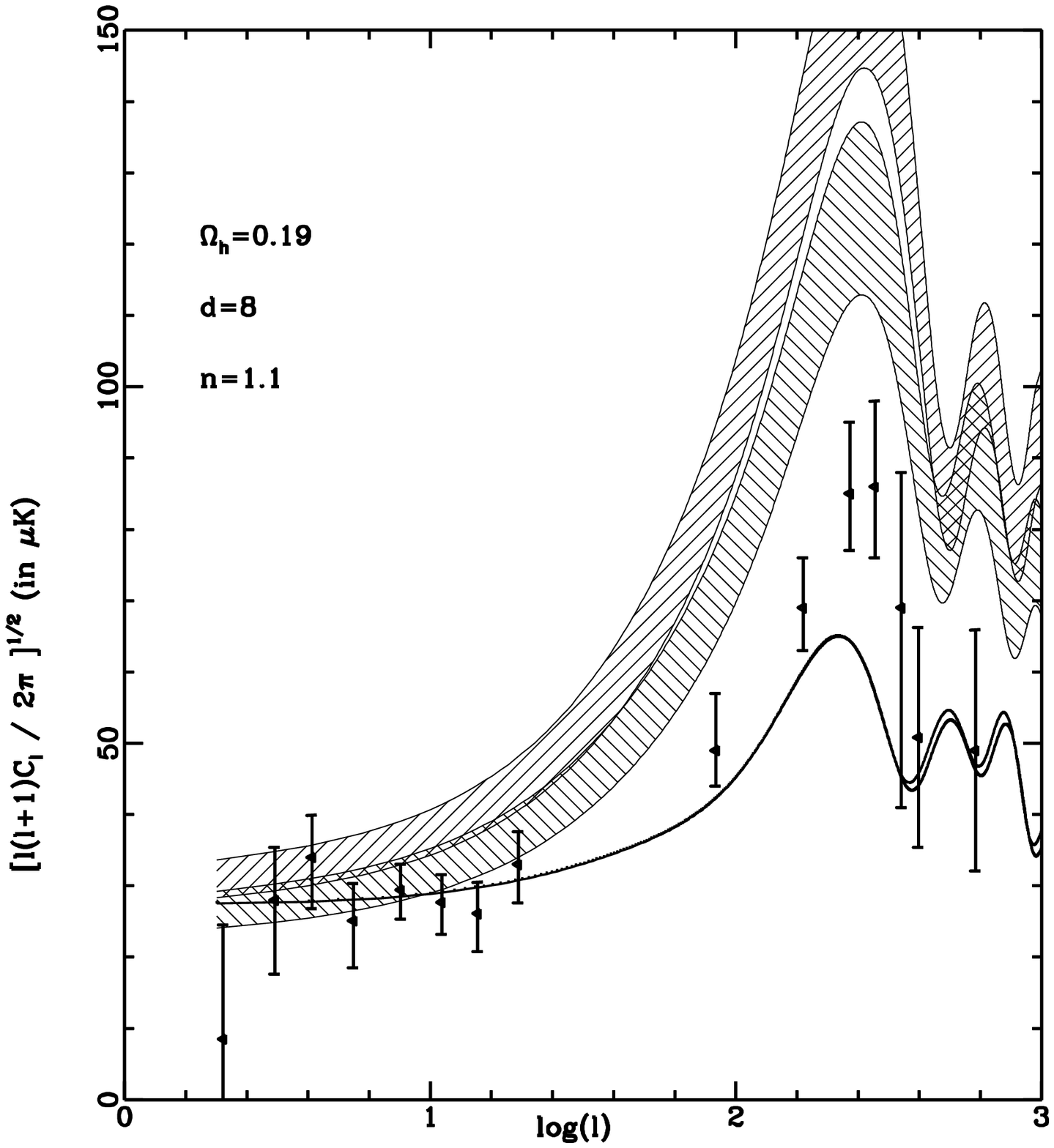}}}
\caption{\it volatile spectra compared to the data. $z_{der}=1250$.}
\end{figure}

\begin{figure}
\centering
\centerline{\mbox{\epsfysize=10.0truecm\epsffile{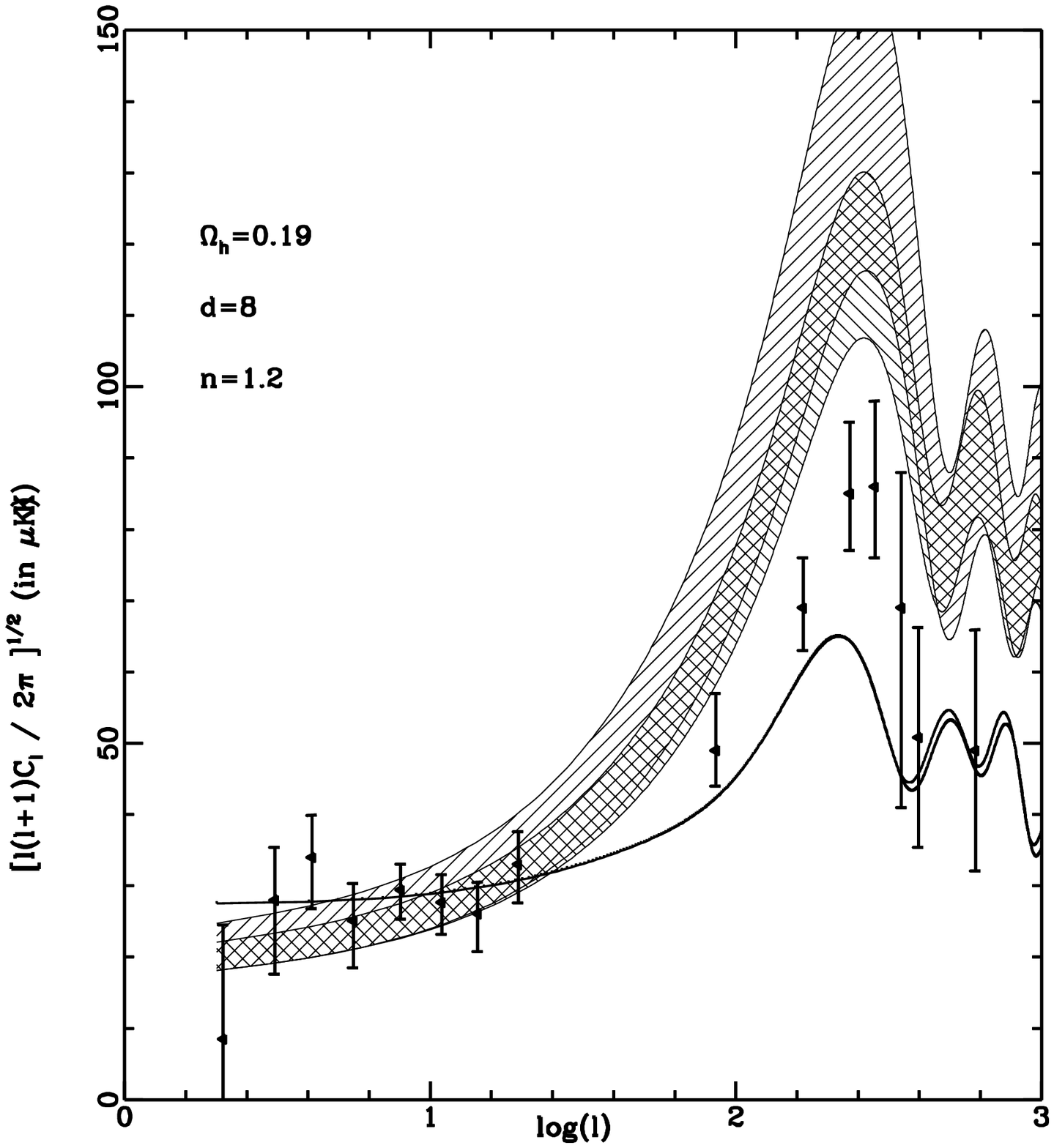}}}
\caption{\it volatile spectra compared to the data. $z_{der}=1250$.}
\end{figure}

\begin{figure}
\centering
\centerline{\mbox{\epsfysize=10.0truecm\epsffile{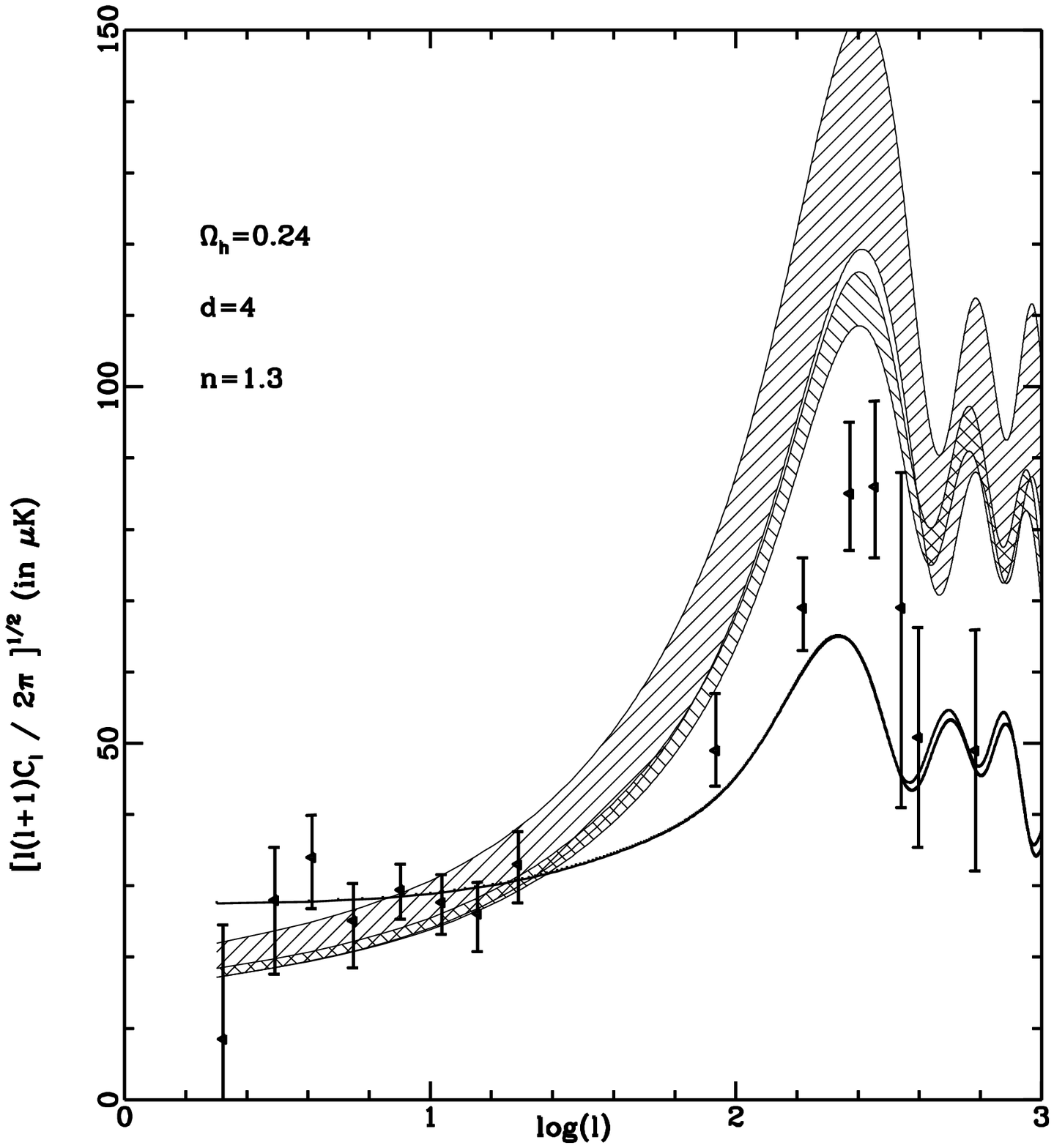}}}
\caption{\it volatile spectra compared to the data.  $z_{der}=2500$.}
\end{figure}

\begin{figure}
\centering
\centerline{\mbox{\epsfysize=10.0truecm\epsffile{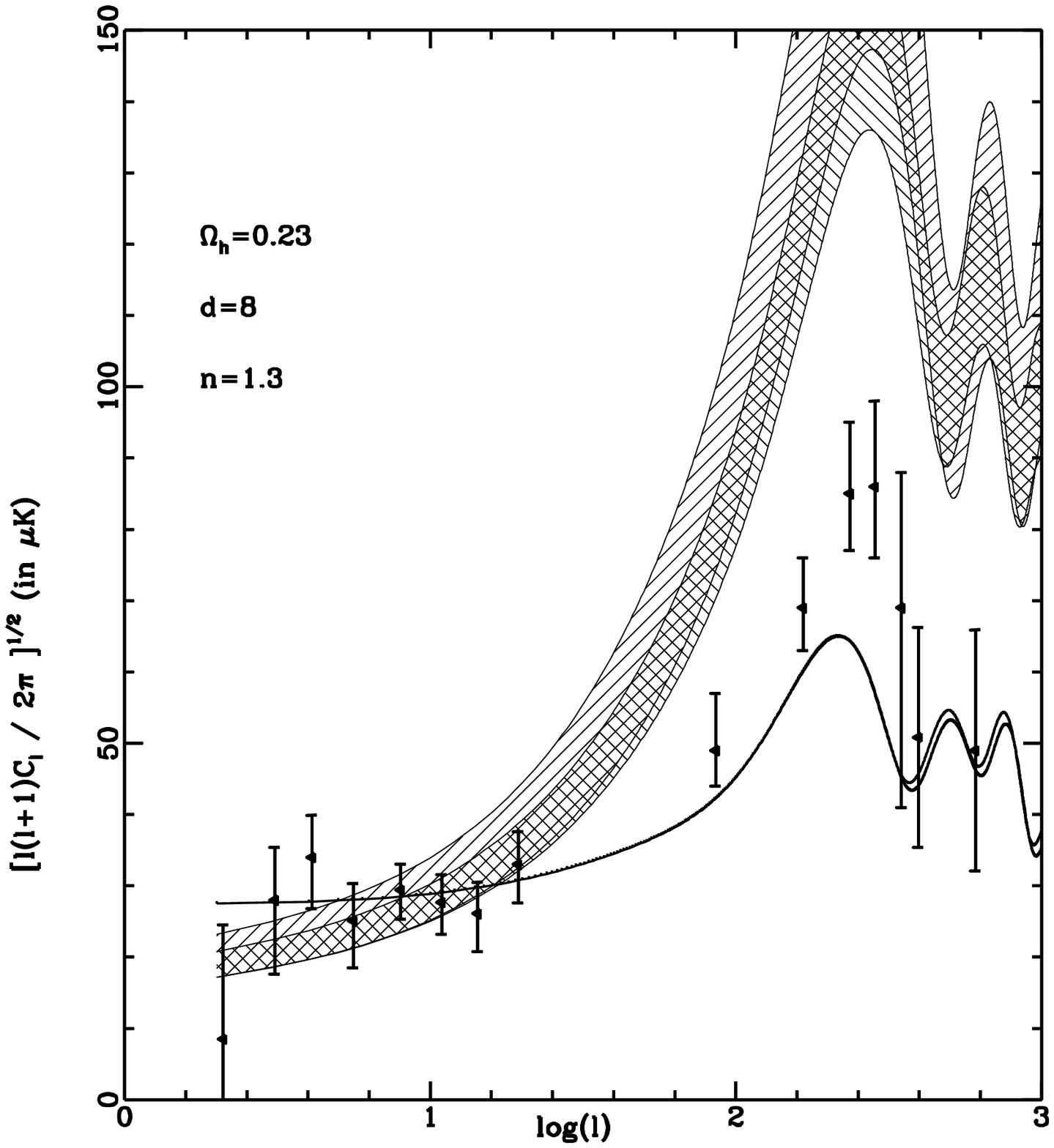}}}
\caption{\it volatile spectra compared to the data. $z_{der}=1250$.}

\end{figure}

While in fig.~5 we plot most data are 
available,
in figs.6--15 we compare models with  data from COBE (Tegmark 1996)
Saskatoon (Netterfiled et al. 1997) and CAT (Scott et al. 1996) only.

Figs.~6--15 show a systematic trend: for a given large $l$ value, $C_l$ 
increases with both $n$ and $d=10^4/z_{der}$.
On the contrary, for a given large $k$ value, the matter fluctuation 
spectrum $P(k)$ increases with $n$ but is damped for large $d$,
so that these to effects tends to compensate.

This is one of the reasons why LSS constraints can be compatible with 
 $n$ as high as 1.4.
On the contrary, figs.14--15 show that CBR spectra already disfavour $n=1.3$ if 
$d \magcir  4$ ($z_{der} \mincir 2500$) is considered, no matter the value 
of $\Omega_b$. Volatile models with $n > 1.2$ are largely out of the 
errorbars, and should  be considered as scarcely viable.
Nevertheless, even for $n=1.2$, volatile models
allow a higher first doppler peak without raising the 
following ones, and therefore  fit the data better than neutrino models. 
Just as large $n$, also large $d$ causes conflict with data, by itself.
For example, fig.~10 show that
models with  $d=16$ are disfavoured, 
even with low $\Omega_h$ and $n=1.1$.

As pointed out in section 2, volatile models require a  sterile component 
whose energy density is proportional to $\Omega_h~d$.
Its effective number of degrees of freedom is linked to the equivalence 
redshift, which in turn affects both the shape parameter $\Gamma$ and 
the height of the first doppler peak.
Dodelson et al. (1994) considered the matter power spectrum
in the case of a $\tau$--neutrino decay ($\tau$CDM model), and found that
even in that case the effective number of degrees of freedom $g_{\nu,eff}$
is bigger that in standard CDM. 
They outlined that, in order to lower  $\Gamma$ at least down to 0.3 
(Peacock \& Dodds 1994),
in a $h=0.5$ universe  with $n=1$, an equivalent number of massless 
neutrinos as high as 16 is needed.  
White et al. (1995), who also consider $\tau$CDM models but with a 
lighter neutrino, also pointed out that the predicted $\Gamma$ of these models
is lower due to the high  $g_{\nu,eff}$, and show that a lower $\Gamma$ 
implies a higher first doppler peak. 
Their work, however, is only qualitative, and they don't infer any 
restriction in the parameter space using the data.
Lookng at the data,   
we found out that if $n=1.1$, CBR data models with an equivalent 
number of neutrino species  $N_{\nu} \magcir  10$ as in figs.~10,~12
and 15 are disfavoured .
Models like the one shown in fig.~8 ($N_{\nu} \simeq 7$) seem to 
better fit the data, although even lower  $N_{\nu}$ ($\simeq 5$),
as provided by the model in fig.~9, should not be disreguarded.

Keeping to  $n=1$,  LSS already exclude very high $\Omega_h~d$ values,
so that a low $N_{\nu}$ is automatically ensured.
The models shown in figs.~6--7 seem to well fit the data,
with a corresponding  $N_{\nu} = 5-7$.

\section{Conclusions}

In this work we have analized mixed models  from the point of view of
both LSS and CMB predictions.
We considered different hot dark matter components: the standard neutrino
case and the volatile case in which particles come from the decay of
heavier ones.

First we tested the mixed models on  available  LSS  data
requiring fair predictions for
$\sigma_8$, $\Gamma$, DLAS and $N_{cl}$.
This analysis 
shows that it must be
$\Omega_h 
\mincir 0.3$.
This comes as no surprise, as mixed models with greater $\Omega_h$
have not been considered since long. The new result is that taking
$n$ up to 1.4 does not ease the problems previously found for large 
$\Omega_h$.

On the contrary, volatile models together with $n > 1$ significantly
widen the parameter space in the low $z_{der}$ direction and
viable models even with $z_{der} \sim 600$ can be found.
In fact, as far as $P(k)$ is concerned, we found a nearly
degenerate behaviour
of the parameters $n$ and $z_{der}$,
as the
damping on the high $k$ values 
due to low
$z_{der}$ can be 
compensated by 
high $n$.

CBR data, apparently,
%
break the degeneracy.
In section~3 we  have shown that the CBR spectrum of volatile models 
is significantly different from 
standard CDM 
and also from 
 neutrino models usually considered.
In fact
SMLC 
and late $z_{der}$ volatiles
cause a late 
$z_{eq}$
and, 
henceforth,
a higher first 
doppler peak.
Minor effects are caused by the typical 
momentum
 distribution 
of volatiles.
These effects amounts to 2 \% at most in the $C_l$ spectrum and 
only an accurate analysis of the results of future satellites, as
MAP and Planck,
could allow to detect it.

CBR spectra of volatile models were then
compared 
with
available data from
different experiments, 
namely those from
COBE,
Saskatoon and CAT experiments.
Data on CBR spectrum at large $l$ imply that temperature
fluctuations $\Delta T/T \sim 10^{-5}/l$ are appreciated.
Therefore, measures of 
the CBR spectrum, for high $l$ values,  still
need to be treated with some reserve.
It seems however clear that recent
observations
tend to indicate a doppler peak higher than expected both for pure
CDM and for mixed models with early derelativization, such as most
neutrino models. 
Taking $n>1$ and/or late
derelativization raises the doppler peak and affects
the CBR spectrum at high $l$. The first question we tried to answer
is how far we can 
and have to
go from pure CDM and $n=1$ to meet current large $l$ data. 


We found that
volatile models could 
cure this discrepancy, while ensuring a viable 
scenario for  structure formation.

In turn, 
large $l$
data 
imply restrictions in the parameter space, 
complementary to the ones derived
from LSS
while a fit of such data requires only a slight departure from pure CDM
and $n=1$.
allows us to say that mixed models are in  very good shape.
 For example, fig.s 8-9 show the $C_l$ behaviour
for $n=1.1$ and HDM ranging from 11$\, \%$ to 16 \%.
Such models provide excellent fits to current data and, as
explained in BP, 
are also in agreement with LSS.

Other models, for larger $n$ and $\Omega_h$ or lower $z_{der}$,
show only a marginal fit with current
observations. Hopefully, future data on high $l$'s will
be more restrictive and allow safer constraints.
At present, such models cannot be ruled out, although they
are more discrepant from pure CDM and $n=1$ than high $l$ data 
require.

In our opinion, however, CBR data can  already be said to 
exclude a number of models which fitted LSS data. In general,
models with $n > 1.3$ and $z_{der} < 1000$ seem out of the range
of reasonable expectations.

Altogether, three kinds of departures from CDM and Zel'dovich were considered
in this work: large $\Omega_h$, low $z_{der}$ and $n>1$. Large (but allowed)
$\Omega_h$ values, by themselves, do not ease the agreement of
models with high $l$ data. Taking $n>1$ eases the agreement of models
with data for $l \sim 200$, as is expected, but seems to rise the
angular spectrum above data for greater $l$'s. Taking low $z_{der}$,
instead, raises the doppler peak, but does not spoil the agreement
with greater $l$ data. Current data, therefore,
seems to support models with a limited amount of HDM or volatile
materials, possibly in association with $n$ slightly above unity,
to compensate some effects on LSS.

Note that the analysis of this work is carried out
keeping $h=0.5$, allowing for no cosmological constant, and
constraining the total density to be critical. E.g., raising $h$
would probably allow and require a stronger deviation from pure CDM
and $n=1$. We plan to widen our analysis of the parameter
space in the near future, also in connection with the
expected arrival of fresh observational data on the CBR spectrum.


\section*{Acknowledgments}

E.P. wishes to thank the University of Milan for its hospitality during
the preparation of this work.

\vfill\eject

\centerline{\bf APPENDIX A}
\vglue 0.8truecm

\section{ Inflationary models yielding $n>1$ }

This section is a quick review of results in the literature,
aiming to show that there is a wide class of inflationary
models which predict $n > 1$, but $\mincir 1.4$.

During inflation, quantum
fluctuations of the {\sl inflaton} field $\varphi$
on the event horizon give rise to density fluctuations. Their amplitude
and power spectrum are related to the Hubble parameter $H$ during inflation 
and to the speed $\dot \varphi$ of the {\sl slow--rolling--down} process. 
The critical quantity is the ratio $W(k) = H^2/\dot \varphi$, where $H$ and 
$\dot \varphi$ are taken when the scale $2\pi/k$ is the event 
horizon. It can be shown that
$$
n = 1 + 2 {d(\log\, W) \over d(\log\, k)}
\eqno (a.1)
$$
and, if $W$ (slowly) decreases with time, we have the standard case
of $n$ (slightly) below unity.
Such decrease occurs if the downhill motion of
$\varphi$ is accelerated and an opposite behaviour occurs if $\dot \varphi$
decreases while approaching a minimum. The basic reason why a potential
yielding such a behaviour seems unappealing, is that the very last stages of
inflation should rather see a significant $\varphi$--field acceleration,
ending up into a regime of damped oscillations around the true vacuum, when 
reheating occurs.

However, the usual perspective can be reversed if
the reheating does not arise when an initially smooth acceleration finally
grows faster and faster, but is triggered by an abrupt
first order phase transition, perhaps due to the
break of the GUT symmetry. Before it and since
the Planck time, most energy resided in potential
terms, so granting a vacuum--dominated expansion. This picture of
the early cosmic expansion is the so--called {\sl hybrid 
inflation}, initially proposed by Linde (1991a).

A toy--model to realize such scenario (Linde 1991b, 1994) is obtainable from
the potential
$$
V(\varphi,\chi) = (\mu^2 - \lambda \chi^2)^2 + 2g^2 \varphi^2 \chi^2~,
+ m^2 \varphi^2
\eqno (a.2)
$$
which depends on the scalar fields $\varphi$ and $\chi$, that we shall
see to evolve slowly and fastly, respectively. If the {\it slow}
field is embedded in mass terms, the potential reads
$$
V(\chi) = M^2 \chi^2 + \lambda \chi^4 + \Lambda^4
\eqno (a.3)
$$
where
$$
\Lambda^4 = \mu^4 + m^2 \varphi^2 ~~~~~{\rm and} ~~~~~~
M^2 = 2(g^2 \varphi^2 - \lambda \mu^2) ~.
\eqno (a.4)
$$
Eq.~(a.3) shows that $V$ has a minimum at $\chi = 0 $, provided that $M^2 > 0$.
If  $M^2 < 0$, instead, the minimum is for $\bar \chi$$= 
\sqrt{-M^2/2\lambda}$, yielding $\mu$ when $\varphi = 0$.

Large $\varphi$ values therefore require that $\chi$ vanishes and then
the potential
$$
V(\varphi,0) = \mu^4 + m^2 \varphi^2
\eqno (a.5)
$$
gives a Planck--time inflation, as $V$ goes from an initial value 
$\sim t_{Pl}^{-4}$ to a value $\sim \mu^4$. The downhill motion of $\varphi$
will decelerate as soon as the second term at the $r.h.s.$
of eq.~(a.5) becomes negligible, in respect to $\mu^4$, which
acts as a cosmological constant. This regime breaks down
when the critical value $\varphi_c = \sqrt{\lambda} \mu/g$ is attained.
Then $M^2$ changes sign and the configuration $\chi = 0$ becomes
unstable. We have then a transition to the true vacuum configuration 
$\bar \chi$, which reheats (or heats) the Universe.

There are several constraints to the above picture, in order that
at least 60 e--foldings occur with $\varphi > \varphi_c$
and fluctuations have a fair amplitude. They are
discussed in several papers (see, $e.g.$, Copeland et al. 1994, and 
references therein) and cause the restriction $n \mincir 1.4$. 

It is fair to outline that {\sl hybrid} inflation is not just one
of the many possible variations on the inflationary theme. In spite of
the apparent complication of the above scheme, it is an intrinsically simple
picture and one of the few
patterns which can allow to recover a joint particle--astrophysical
picture of the very early Universe.

\end{document}